\documentclass[a4paper]{article}

\usepackage{INTERSPEECH2018}
\usepackage{setspace}
\title{Leveraging native language information for improved accented speech recognition}
\name{Shahram Ghorbani$^1$, John H.L. Hansen$^2$ \thanks{This project was funded by AFRL under contract FA8750-15-1-0205
and partially by the University of Texas at Dallas from the Distinguished University Chair in Telecommunications Engineering held by J. H. L. Hansen.}}
\address{Center for Robust Speech Systems (CRSS), 
University of Texas at Dallas, Richardson, TX 75080}
\email{shahram.ghorbani@utdallas.edu, john.hansen@utdallas.edu}

\begin{document}
\maketitle

\begin{abstract}
Recognition of accented speech is a long-standing challenge for automatic speech recognition (ASR) systems, given the increasing worldwide population of bi-lingual speakers with English as their second language. If we consider foreign-accented speech as an interpolation of the native language (L1) and English (L2), using a model that can simultaneously address both languages would perform better at the acoustic level for accented speech. In this study, we explore how an end-to-end recurrent neural network (RNN) trained system with English and native languages (Spanish and Indian languages) could leverage data of native languages to improve performance for accented English speech. To this end, we examine pre-training with native languages, as well as multi-task learning (MTL) in which the main task is trained with native English and the secondary task is trained with Spanish or Indian Languages. We show that the proposed MTL model performs better than the pre-training approach and outperforms a baseline model trained simply with English data. We suggest a new setting for MTL in which the secondary task is trained with both English and the native language, using the same output set. This proposed scenario yields better performance with +11.95\% and +17.55\% character error rate gains over baseline for Hispanic and Indian accents, respectively.
\end{abstract}

\section{Introduction}

In recent years, deep neural network (DNN) models have replaced conventional acoustic models in many ASR applications and made significant improvements in ASR system performance. One successful network architecture used for speech processing is RNN. RNNs perform well as acoustic models in hybrid DNN-HMM systems \cite{graves2013hybrid,sak2014,hsiao2014improving}, which confirm their capabilities to reflect long-range dependencies of input sequences within end-to-end models  \cite{graves2013,deepspeech2,mirsamadi2017multi}. The output of such networks can be phonemes or graphemes, where for large data sets allow performance of phoneme models to match that of grapheme models \cite{accent2017}.

Current ASR systems perform well for close-talk speech recognition scenarios and perform better than humans in some benchmarks \cite{best1,best2}. However, real scenarios are more complex than those benchmarks, and ASR systems should overcome several challenges to be able to reach human performance. One challenging real aspect is sustained performance for accented speech. Increasing worldwide communication has expanded the number of second language learners that, due to the impact of their first language, speak second languages with varying degrees of accent. The extent of degradation in recognition accuracy for accented speech is dependent on the level of accent and the difference between native and second language. We will show in our experiments how an acoustic model trained with native English performs for different accents.

One solution for training an effective accent robust acoustic model is to collect sufficient data for alternate accents and train a multi-accent model \cite{sainath2017no,accent2017,li2017multi,ghorbani2018advancing}. Kanishka et al. in \cite{accent2017} used a hierarchical grapheme-based model trained in a multitask context, where the model was trained for 4 different accents using two types of cost functions, phoneme-based, and grapheme-based, and obtained an improved multi-accent model. Another way to train a multi-accent model is to feed bottleneck or accent-specific features to the acoustic model as auxiliary information \cite{chineseaccent2016,li2017multi}. In cases where sufficient data is not available for each accent, one can adapt a pre-trained model to the target accent. It is important to avoid overfitting the model which can be accomplished by selecting a subset of layers to tune, or a generalization constraint such the KL-divergence \cite{huang2015regularized}.
\begin{figure*}[ht!]
  \centering
  \includegraphics[width=1\textwidth,height=2.9cm]{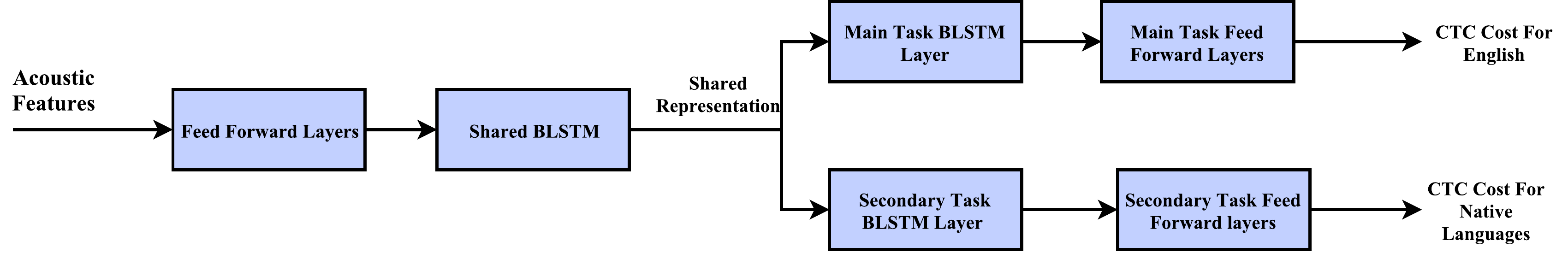}
  \caption{The proposed multitask architecture to leverage native language data.}
  \label{fig:framework}
\end{figure*}

Collecting sufficient accented data to train accent specific models from scratch may not be feasible due to the diverse number of possible accents and limited available population of speakers within an accent. Many previous studies have shown that native language L1 affects speaker traits of their secondary language (e.g. English) \cite{multi1,multi2,multi3}. Therefore, it is reasonable to use data from both L1 and L2 to train a model to be more tolerant of pronunciation variants of accented speech. In addition, it is easier to collect data from native speakers than accented speakers, and data for most secondary spoken languages are more easily available. Having a model which exploits data of native non-English languages to train a robust acoustic model for foreign-accented speech, would result in improved ASR systems for many accents without collecting new accented data. In this study, we investigate how to use the data of L1 and L2 languages to train an end-to-end model in a multitask and pre-training settings to investigate the effect of bilingual training on non-native speech recognition. Due to this two-step training, it is expected that pre-training will have only small model benefits. In the multitask setting, the secondary task is trained with Spanish and Indian languages and the primary task is trained with native English data, as shown in Fig \ref{fig:framework}. Training the shared portion of the network to minimize the costs of both languages potentially results in a model that performs well for accented speech corresponding to native languages. In a different novel setting, we train the secondary task with both L1 and L2 yielding a more advanced model and significantly improves recognition performance for accented speech.

\section{RNN-CTC acoustic modeling}

An acoustic model receives a sequence of input features $ \textbf{X}=\{x_1, x_2,..., x_T\}$ and maps them to a sequence of distributions over output characters $\textbf{Y}=\{y_1,y_2,..., y_T\}$. Generally, a recurrent neural network estimates a per-frame un-normalized distribution over the output symbols using:

\begin{equation}
    \begin{aligned}
    \textbf{h}_t = \mathcal{H}(\textbf{W}_{ih}x_t + \textbf{W}_{hh}\textbf{h}_{t-1}+\textbf{b}_h),\\
    \textbf{y}_t = \textbf{W}_{ho}\textbf{h}_{t}+\textbf{b}_o,
    \end{aligned}
\end{equation}

\noindent
where \textbf{W} and \textbf{b} are the weights and biases of the network, respectively, $\mathcal{H}$ is the activation of the layers and $y_t$ is the output of the network corresponding to input $x_t$.

Long Short-Term Memory (LSTM) networks use gated memory cells to store information that enables them to exploit long-range context, and for some scenarios could outperform conventional recurrent neural networks for speech recognition\cite{sak2015fast}. In \cite{BLSTM}, it was shown that using input features in both directions  (forward and backward) improves performance of the acoustic model compared to a unidirectional trained model. In this study, we also employ bidirectional LSTM cells in the recurrent layers. 

To train the LSTM-CTC network, we need alignments between input and output sequences which for most speech corpora are unknown. One approach to address this problem is Connectionist Temporal Classification (CTC) which enables us to train a model that maps the input sequence (frames) to an output sequence (graphemes) without requiring any strict alignment \cite{ctc}. 

In our end-to-end architecture, the size of the last layer is equal to $| \textbf{S}|$, where \textbf{S}=\{characters\,of\,the\,language, ‘blank’, space, noise\}. Here, 'blank' is a special character used by CTC for calculating the cost of the output and also by decoder to output the final sequence. Given a sequence of feature vectors  $\textbf{X}$, the output of the last layer of the network is submitted to a softmax layer to produce the probability of the members of $S$ for any input frame:
\begin{equation}
Pr(k,t|\textbf{X}) = \frac{exp(y_k^t)}{\sum_{j=1}^{|S|} exp(y_j^t)},   
\end{equation}

\noindent
where  $y_k^t$  is the probability of emitting the \textit{k}th member of $S$ for the given input $x_t$. We can consider the result of the network for a given sequence $X$ as a matrix $\textbf{O}$ of size $|\textbf{S}|* T$ from which by choosing one element of each column we obtain a length $T$ output sequence where its  probability is $ Pr(a|\textbf{X}) = \prod_{t=1}^{T} \textbf{O}(a(t),t)$, and the CTC objective is to maximize the probabilities of such sequences that correspond to the target labels:

\begin{equation}
\theta = argmax\sum_{a\in \mathcal{A}} \prod_{t=1}^{T} \textbf{O}(a(t),t).
\end{equation}

Here, $\mathcal{A}$ is the set of all alignments related to the target sequence and $\theta$ represents the parameters of the neural network. Next, given a new input sequence to the trained network, the decoder should find the most probable output transcription. The decoding could be accomplished by simply choosing the most probable output from each column of $\textbf{O}$. This scenario is often referred to as "best path decoding" or, to get a more accurate sequence we could employ a version of "beam search decoding" \cite{graves2013}. In both approaches, the decoder should consider removing any blank and label repetitions from the final transcription.  

\begin{table*}[ht!]\caption{Corpus of UT-CRSS-4EnglishAccent with the native and non-native accents US, Australian, Hispanic (Mexican Spanish accent of English) and Indian.}\label{tab:datastat}
\centering
\scalebox{1}{
\begin{tabular}{l p{.7cm}p{0.7cm}p{0.7cm} p{.7cm}p{0.7cm}p{0.7cm}p{.7cm}p{0.7cm}p{0.7cm}p{.7cm}p{0.7cm}p{0.7cm}}
\hline
& \multicolumn{3}{c}{\textbf{US}} & \multicolumn{3}{c}{\textbf{Australian}} &  \multicolumn{3}{c}{\textbf{Hispanic}}  &  \multicolumn{3}{c}{\textbf{Indian}} \\
\textbf{ } & \textbf{Train} & \textbf{Dev} & \textbf{Test} & \textbf{Train} & \textbf{Dev} & \textbf{Test} & \textbf{Train} & \textbf{Dev} & \textbf{Test}  & \textbf{Train} & \textbf{Dev} & \textbf{Test} \\ \hline \hline
\textbf{\#speakers} & 64 & 23 & 28 & 60 & 15 & 16 & 64 & 22 & 26 & 64 & 22 & 26  \\ 
\textbf{male/female} & 31/33 & 11/12 & 13/15 & 28/32 & 7/8 & 7/9 & 34/30 & 12/10 & 14/12 & 34/30 & 12/10 & 14/12  \\
\textbf{length [h]} & 27.1 & 4.8 & 5.4 & 22.2 & 3.4 & 3.5 & 28.8 & 6 & 6.5 & 28.35 & 5.6 & 6.2 \\ 
\textbf{\#read utts} & 16k & 1.9k &2.4k  & 14.5k & 1.6k & 1.6k & 16k & 2.5k & 2.5  & 16k & 2k & 2.4k \\
\textbf{\#spont utts} & 950 & 340 &400  & 868 & 215 & 225 & 1k & 380 & 387 & 1k & 400 & 380 \\
\textbf{\#total utts} & 17k & 2.3k & 2.8k & 15.4k & 1.8k & 1.8k & 17k & 2.8k & 2.9k & 17k & 2.4k & 2.8k \\\hline
\end{tabular}
}
\end{table*}

\section{Leveraging native language data}
In this section, we describe our proposed training strategies for improved recognition of accented speech. In this scenario, we only have native English data, and data from native language speakers corresponding to the target English accent (i.e., Spanish data for Hispanic English accent and Indian languages data for Indian English accent). A resembling problem in speech recognition is improving ASR modeling for low resources languages using available data of other languages. A conventional approach in the multi-lingual domain is to train a network to provide a multi-lingual generic representation from which we could train a simpler model with limited language data. However, our problem is slightly different, where the data of target accents is not available but we instead have data of the target language in a different accent (i.e., the data for native English which is adequate to train an ASR model).

The first proposed method to exploit data of L1 language employs pre-training approaches. First, we train a neural network model with native language from which we keep all but the last two layers as a pre-trained model for next step. On top of the pre-trained network, we build a new LSTM and fully connected layers trained from scratch with native English data. We find that this setting of pre-training is optimal for our problem.

One drawback of this pre-training approach is that by adapting the model to US English, much of the learned information from the native languages would be forgotten. This two-step training would not leverage the available information efficiently. To address this problem, we propose a multitask learning as shown in Figure \ref{fig:framework} where the primary task (Task1) is trained with English data and the secondary task (Task2) uses data from the target native language. Grapheme-based CTC losses are calculated for both languages (TC1 and TC2) and backpropagated through each task and combined with a combination factor $\lambda$ to train the shared parts:
\begin{equation}
    Total\_Cost = (1-\lambda)*TC1 + \lambda*TC2.
\end{equation}

In this scenario, we have all native language data available in the training which by choosing the best value of $\lambda$ the model exploits data of native language efficiently in favor of improved ASR model for accented speech. Training the shared part of the network with the combined cost results in a network that represents similar phonemes of the two languages with a similar vector. Therefore, given accented speech corresponding to the L1 language, the multi-task trained model would provide a representation that the main task has received in training steps and is expected to be more successful than a native English only trained model.  

Training an LSTM-CTC model with utterances from the two languages using the same character set would result in a network that maps all similar phonemes of both languages to the same representation. Such a network might not perform well for each language individually, but could be exploited to provide a bilingual representation. To this end, we also propose a new setting for multitask learning in which the secondary task recognizes both L1 and L2 with the same character set. This setting, even more, drives the network to better represent close phonemes of L1 and L2 with a similar representation, yielding an improved performance of the main task in recognizing L1 accented English.

\section{System setup and data}
For training and evaluating the model across accents, the UT-CRSS-4EnglishAccent corpus is used. This corpus of 420 speakers was collected at CRSS UTDallas and consists of four major English accents: US (native), Hispanic, Indian and Australian. The data for each accent consists of about 100 speakers balanced for gender and age, with session content consisting of read and spontaneous speech. In this study, we use the US and Australian portions of the corpus to train the "English" main task, and the Hispanic and Indian accent portions as accented English test sets. Table \ref{tab:datastat} presents some statistical information about the train, development and test sets of the four English accents.

To obtain training data for our secondary task, we need native speech in Spanish and Indian languages. In the case of Spanish, the CIEMPIESS corpus is used \cite{spanishcorpus}. This corpus is collected from broadcast news of Mexican Spanish and provides 17h of training data. In the case of Indian languages, since many languages are spoken in India it is not possible to associate the Indian accent to a single L1 language. Therefore, we have chosen the six most spoken languages in India; Hindi, Kannada, Gujarati, Marathi, Tamil, and Telugu. For each language 5h of training data is selected from the IndicTTS \cite{IndianCorpus} corpus, resulting in a total of 30h of training data. In order to obtain a same character set for all languages, using \cite{SILPA} all scripts are converted to English characters. having a common output set based on English ensures a common partitioned acoustic space for our multitask system.
 
We extract 26 dim Mel filterbank coefficients for each 25ms frame with an overlap rate of 10ms. We expand each frame by stacking 4 frames on each side, then frames are decimated by skipping 2 frames per each frame for processing. The skip process is described in more detail in \cite{sak2015fast}.

The baseline model has the same architecture as the main branch of our multitask model (see Figure \ref{fig:framework}), consisting of two bidirectional LSTM layers each with 300 cells followed and preceded by two feed forward layers each with 500 neurons. The activation of the last layer is a softmax that outputs the probability of each character plus 3 outputs for $‘blank’$, space and noise. For English as well as Indian Languages, there are 29 outputs, and for Spanish with the T22 \cite{spanishcorpus} phoneme set, there are 25. All model parameters are optimized using an Adam Optimizer with an initial learning rate of 0.001. All model weights are initialized with a random value from a normal distribution with standard deviation of 0.04. For model training, mini batches of size 30 utterances are employed. We engage early-stopping for training by monitoring the performance on a held-out validation set during training. We employ beam search decoding with a beam-width of 100 and note that no language model or lexicon information is used. 

For bilingual multitask learning as shown in Figure \ref{fig:framework} a single LSTM layer preceded by two fully connected layers are shared and trained for the two tasks where the combination factor $\lambda$ is $0.3$. For the secondary task on top of the shared layers, two types of architecture are examined; (i) a large architecture with one LSTM layer followed by two feed forward layers, and (ii) a small architecture that simply has the feedforward layers.

\section{Results and discussion}

\subsection{Baseline Model}
\label{sssec:subsubhead}
This section presents results to assess how much accent mismatch between training and testing impacts the performance of the acoustic model. To ensure sufficient data to train the baseline model, training data from US and Australian portions of UT-CRSS-4EnglishAccent corpus are combined, and the trained model is tested on US, Hispanic and Indian accents.
\begin{table}[!th]
\caption{\label{tab:mono} {\it CER on English with native and non native accents using a baseline English acoustic model trained with US English (US) and Australian English (AUS). }}

\begin{center}
\begin{tabular}{ p{3cm} c c c }
    \hline
        Task1 Train Data &  US  & HIS Acct & Indian Acct \\
      \hline
     US, AUS & 15.2 & 18.4 & 31.9 \\ 
	  \hline
\end{tabular}
\end{center}
\end{table}

Table \ref{tab:mono} shows resulting CERs for the test datasets. The baseline model obtained a CER of 15.23\% for the US portion of native English test set, but performance drops when presented with Hispanic accented English from $15.23\% \to 18.41\%$. A significant loss in performance also occurs for Indian accented English, with CER increasing from $15.23\% \to 31.86\%$.  Since the acoustic recording conditions were held constant for all accents, we suggest that the performance degradation is due to a phonetic mismatch between native and non-native speakers.

\subsection{Bilingual Models}
In this section, we consider the proposed multitask and pre-trained model solutions to assess the potential benefits of these architectures where information from native languages are leveraged to improve model performance for accented speech. Pre-training the shared portion with native languages provides a good initial model for the main task. However, since the training data for the tuning step are not accented, the model forgets most of the information that is learned from native languages and could not significantly benefit the combined model (Row 2 of Table \ref{tab:bilingual}). 

For the MTL approach, using a large network for the secondary part provides more parameters to learn in Task2 and would therefore have less impact on the shared portion. Alternatively, using a small network for Task2 might bias the shared portion towards the native languages and hurts the main task performance. 

Rows 3 and 4 of Table \ref{tab:bilingual} show performance of our proposed multitask approach for large and small architectures for the secondary task. We find that our model with a multitask setting always outperforms the baseline and improves CER from $18.4\% \to 17.7\%$ for Hispanic and $31.9\% \to 29.5\%$ for Indian accents, and shows that our approach has trained the shared portion of the network in a positive direction. We see better performance for the small network of Task2 as well,  where CER is improved a relative +6.2\% for Hispanic and +3.0\% for Indian accents compared to the large network, showing that the small configuration provides a better representation.

We also examine this proposed architecture where the secondary task is trained with both languages; English and Indian languages for Indian accent, and English and Spanish for Hispanic accent. Two large and small networks for Task2 are evaluated with performance for accented test sets shown in rows 5 and 6 in Table \ref{tab:bilingual}, respectively. Bilingual training of the large secondary task does not significantly outperform the previous best model (Row 5 vs. row 4 of Table \ref{tab:bilingual}). However, for the smaller network case in Task2, model performance is significantly better and achieves a relative gain of +11.95\% for Hispanic and +17.55\% for Indian accent vs. baseline.

\begin{table}[!th]
\caption{\label{tab:bilingual} {\it CER on English with non-native accents for proposed bilingual multitask model trained with either English and Indian Languages (InL), or English and Spanish (SP). }}
\begin{center}
    
\begin{tabular}{ c c c c c  }
    \hline
       Model & Task2 Size &   HIS Acct & Indian Acct \\
      \hline
      Single Task & None & 18.4 & 31.9 \\
      
       \hline
       Pre-training  & None& 17.3 & 30.4 \\
       \hline
       
      MTL, Task2: L1 & Large  & 17.7  & 29.5 \\
      MTL, Task2: L1 & Small   & 16.6 & 28.6 \\
      \hline
      MTL,Task2: L1,L2& Large &16.6 & 28.4\\
      MTL,Task2: L1,L2& Small &\textbf{16.2} & \textbf{26.3}\\
	  \hline
\end{tabular}
\end{center}
\end{table}
\subsection{Adapting Results}

 To investigate if leveraging native language data provides any additional improvement given accented English data, we adapt the baseline model and the best model for accented data (e.g., the model corresponding to row 6 of Table \ref{tab:bilingual}) with Indian and Hispanic accented English data. Here, we use all 28h of training data available for each of the two accents (Table  \ref{tab:datastat}). Adapting the multitask trained model with accented data significantly improves the model performance, especially for Indian accent which is considerably different than the training data (i.e., Australian and US English). Given the adaptation data, the adapted MTL model still outperforms the single task adapted model, achieving a relative gain of  +12\% for Hispanic and +8.4\% for Indian accent. This improvement suggests that the information learned from native languages, with different speakers from English data, benefits model development as well as generalizes the accent specific model compared to the model trained exclusively with English data.
\begin{table}[!th]
\caption{\label{tab:adaptation} {\it CER on English with non-native accents for adapted version of the baseline model and the best multi-task architecture (e.g., the model corresponding to row 6 of Table \ref{tab:bilingual}).  }}
\begin{center}
\begin{tabular}{  c  c c }
    \hline
       Model &   HIS Acct & Indian Acct \\
      \hline
      Adapted Single Task & 15.0 & 14.3 \\
      \hline
      Adapted MTL,Task2: L1,L2    & \textbf{13.2} & \textbf{13.1} \\
	  \hline
\end{tabular}
\end{center}
\end{table}

\subsection{Summary and Conclusions}
In this study, we investigated an improved modeling approach to improve ASR for non-native English. Using an LSTM-CTC model trained with English utterances (US and Australian), we showed that model performance degrades significantly when presented with Hispanic and Indian accented data. To compensate for this degradation, we use data from native Spanish and Indian languages in a pre-training and multitask setting. Pre-training the model with the native language only slightly improves model performance. However, using the native language as the secondary task increases model tolerance to accented speakers. The trained multitask acoustic model outperformed the baseline model, which was exclusively trained with native English data, with a corresponding relative CER decrease of 9.7\% for Hispanic and 10.3\% for Indian accents. Training the secondary task with both native languages (Spanish or Indian languages) and English results in a more accent independent representation and further improves model performance, achieving a relative +11.95\% and +17.55\% CER gain over baseline for Hispanic and Indian accents, respectively. Given accented adaptation data, the adjusted MTL model still outperforms the single task adapted model, achieving a relative CER gain of  +12\% for Hispanic and +8.4\% for Indian accent. These advancements provide promising directions for both maintaining or improving ASR for a wider speaker population based on native and accented subjects.
\newpage
\bibliographystyle{IEEEtran}

\bibliography{mybib}


\end{document}